\documentclass[a4paper,11pt]{article}
\usepackage{pos}

\title{Neutrino Emissions of TXS~0506+056 caused by a Supermassive Binary Black Hole Inspiral?}
 \ShortTitle{Neutrino Emissions of TXS~0506+056 caused by a SMBBH Inspiral?}

\author*[a,b]{Ilja Jaroschewski}
\author[a,b,c]{Julia Becker Tjus}
\author[a,b]{Armin Ghorbanietemad}
\author[d]{Imre Bartos}
\author[a,b,e,f,g]{Emma Kun}
\author[h,i,j]{Peter L. Biermann}

\affiliation[a]{Theoretical Physics IV: Plasma Astroparticle Physics, Faculty for Physics \& Astronomy, Ruhr University Bochum, 44780 Bochum, Germany}

\affiliation[b]{Ruhr Astroparticle And Plasma Physics Center (RAPP Center), Germany}
\affiliation[c]{Department of Space, Earth and Environment, Chalmers University of Technology, 412 96 Gothenburg, Sweden} 

\affiliation[d]{Department of Physics, University of Florida, PO Box 118440, Gainesville, FL 32611-8440, USA}

\affiliation[e]{Astronomical Institute, Faculty for Physics \& Astronomy, Ruhr University Bochum, 44780 Bochum, Germany}

\affiliation[f]{Konkoly Observatory, ELKH Research Centre for Astronomy and Earth Sciences, Konkoly Thege Miklós \'ut 15-17, H-1121 Budapest, Hungary}

\affiliation[g]{CSFK, MTA Centre of Excellence, Konkoly Thege Miklós \'ut 15-17, H-1121 Budapest, Hungary}

\affiliation[h]{MPI for Radioastronomy, 53121 Bonn, Germany}

\affiliation[i]{Department of Physics \& Astronomy, University of Alabama, Tuscaloosa, AL 35487, USA}

\affiliation[j]{Department of Physics \& Astronomy, University of Bonn, 53115 Bonn, Germany}

\emailAdd{ilja.jaroschewski@rub.de}

\abstract{The IceCube neutrino observatory detected two distinct flares of high-energy neutrinos from the direction of the blazar TXS~0506+056:
    a $\sim 300$~TeV single neutrino on September 22, 2017 and a $3.5\sigma$ signature of a dozen TeV neutrinos in 2014/2015.
    In a previous work, it was shown that these two episodes of neutrino emission could be due to an inspiral of a supermassive binary black hole (SMBBH) close to its merger at the core of TXS~0506+056. 
    Such an inspiral can lead to quasi-periodic particle emission due to jet precession 
    close to the final coalescence.
    This model made predictions on when the next neutrino emission episode must occur.
    On September 18, 2022, IceCube detected an additional, $\sim 170$~TeV neutrino in directional coincidence with the blazar TXS 0506+056, being consistent with the model prediction.
    Additionally, in April 2021, the Baikal Collaboration reported the detection of a $224\pm 75$~TeV neutrino, with TXS~0506+056 being in the uncertainty range of the event direction. 
    
    We show that these four distinct flares of neutrino emission from TXS~0506+056 are consistent with a precessing jet scenario, driven by an inspiraling SMBBH.
    Using improved modeling, we are now able to constrain the total mass together with the mass ratio for the binary. 
    We predict when the next neutrino flares from TXS~0506+056 should be happening.
    Finally, we estimate the detection potential of the Laser-interferometer Space Antenna (LISA) for the merger in the future. }

\ConferenceLogo{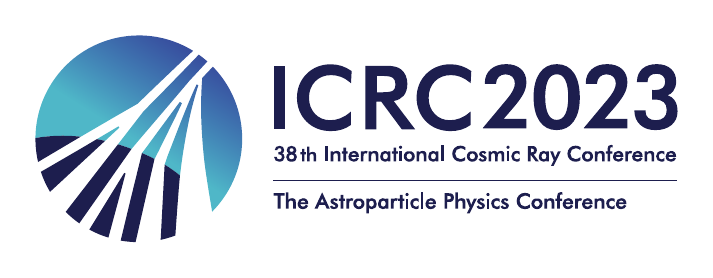}

\FullConference{%
38th International Cosmic Ray Conference (ICRC2023)\\
  26 July - 3 August, 2023\\
  Nagoya, Japan}

\begin{document}
\maketitle

\section{Introduction}

Ever since the detection of a $\sim 300 \,\text{TeV}$ neutrino from the direction of the blazar TXS~0506+056 at a $3\sigma$ level by IceCube, and a coincident detection of a GeV gamma-ray flare by Fermi Large Area Telescope \cite{IceCube2017_2018}, this source became one of the main candidates for the source of cosmic rays (CRs) and neutrinos. 
A blind analysis of $10 \,\text{yr}$ data of IceCube from the same direction revealed another neutrino flare in 2014/2015 with $\sim 10$ neutrinos at $\sim 10 \,\text{TeV}$ at a $3.5\sigma$ level \cite{IceCube2014_2018}.
However, gamma-ray data at that time indicated that the source was in a quiescence mode. 

In September 2022, IceCube reported the detection of another $\sim 170 \,\text{TeV}$ neutrino from the direction of the blazar \cite{Blaufuss2022}, but with no associated gamma-ray detection.
This track-like event has a directional uncertainty of $\sim 3.6^\circ$ ($90\%$ containment) and, with a signalness of $42\%$ of being of astrophysical origin, was classified as a "bronze alert". A reason for this high uncertainty could be that the track skimmed the edge of the IceCube detector and was not fully contained in \cite{Blaufuss2022}. TXS~0506+056 lies in that uncertainty region, with a separation of $\sim 3.06^\circ$ from the best-fit event position.

In addition, the Baikal Collaboration reported the detection of a $224\pm 75 \,\text{TeV}$ neutrino in April 2021 with its uncertainty range covering the direction of TXS~0506+056 \cite{baikal2022}. 
This neutrino is a cascade event and has an uncertainty of $\sim 6^\circ$ ($90\%$ containment), while TXS~0506+056 has a separation of $\sim 5.33^\circ$ from the best-fit event position.
Its signalness was reported as $97.1\%$. 

Though a combined multimessenger modeling of each of these neutrino flares in combination with gamma-ray data is challenging (see e.g.\ \cite{Rodrigues2019, Petropoulou2020}), the distinct arrival times of these flares can be explained by a supermassive binary black hole (SMBBH) close to its merger, which is located at the core of the blazar \cite{britzen2019}.
In its inspiral stage, the emission of gravitational waves (GWs) is the leading mechanism through which the binary loses orbital energy. 
As a consequence, due to spin-orbit coupling, the originally unaligned spins of both supermassive black holes (SMBHs) realign themselves, changing the orientation of the associated jets \citep{G&B2009}.
During this realignment, the jets perform a precessing movement, colliding with the surrounding matter and producing neutrinos due to proton-proton interactions \cite{BeckerTjus2020}.
Is such a precessing jet pointing at Earth, a quasi-periodic neutrino signal will be received, each time the jet finishes one precession.
The time between the signals will shorten will each precession, thus quasi-periodic.
The jet precession model was first developed in \cite{deBruijn2020}.
It was predicted that one flare should occur in the years 2022 and 2023. 
In \cite{BeckerTjus2022}, it was shown that the 2022 IceCube neutrino agrees with this prediction. 
For that, an extension of the model was used \cite{Kun2022}, which considers small mass ratios.

In the following, the jet precession model is extended further to mass ratios up to and including unity by allowing contributions of the second spin and the orbital angular momentum and applied on TXS~0506+056. 
Though the neutrino event detected by the Baikal Collaboration has a high chance of being of astrophysical origin, it remains unclear if it originated from this source, as it has a huge uncertainty range (being a cascade event). 
This is why it is investigated with this updated model, whether all four distinct neutrino flares could originate from TXS~0506+056, if it has an inspiraling SMBBH at its core.
Alternatively, the possibility of only IceCube neutrinos  originating from the source is investigated.

\section{The Jet precession Model}
The model presented here is an extension of the jet precession model first described in \cite{deBruijn2020}.
This model focused on SMBBH mergers with the most common mass ratios between 
$q =1/3$ and $q=1/30$, with $q= m_2/m_1$ and the masses of the binary $m_1 \geq m_2$.
Since the spin magnitude $S_i$ is proportional to $m_i^2$, the contribution of the second spin $S_2$ is ignored in that model, as its magnitude is with $S_2/S_1 \approx q^2$ smaller than $S_1$ \citep{G&B2009}.

\begin{figure}[h]
    \centering
    \includegraphics[angle=0,scale=0.5]{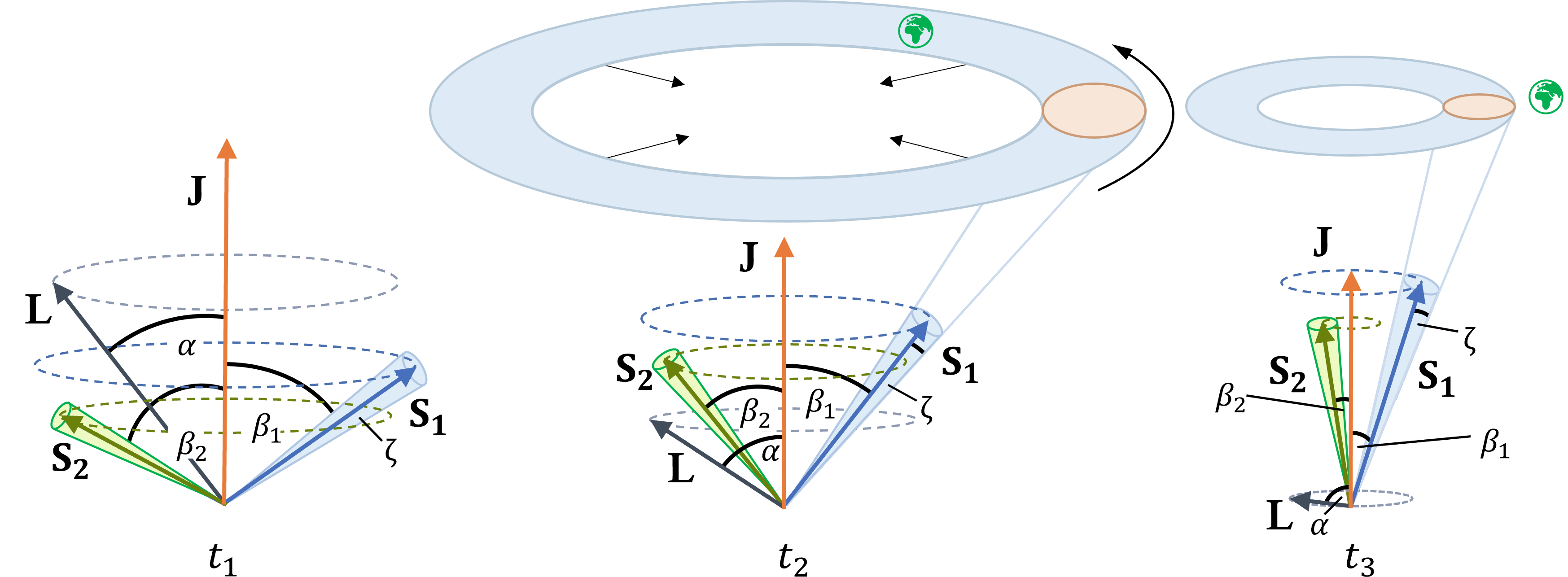}
    \caption{Schematic overview of the jet precession model with two supermassive black holes close to their merger at the center. The jet direction of the heavier black hole $\mathbf{S}_1$, the lighter black hole $\mathbf{S}_2$, the orbital angular momentum $\mathbf{L}$ and the total angular momentum $\mathbf{J}$ are shown. Each time the orange area crosses Earth, a possible neutrino signal can be detected. Figure modified from \cite{deBruijn2020}.}
    \label{fig:Prediction_compare_2Spin}
\end{figure}

In contrast, this new model also considers mass ratios smaller than $1/30$ and higher than $1/3$ up to and including a mass ratio of unity and includes the second spin as well. 
The schematic overview of the jet precession model is shown in Fig.~\ref{fig:Prediction_compare_2Spin}. 
At a time $t_1$, the supermassive binary black hole system enters the inspiral stage.
The spin vectors $\mathbf{S}_1$ and $\mathbf{S}_2$ are most likely unaligned at this time due to the initial random orbital spin orientation of the SMBHs in the galaxy centers of the preceding galaxy merger \cite{Jaroschewski2023}. 
In this stage, the spins couple with the orbit, so that they perform a precessional motion around the orbital angular momentum $\mathbf{L}$ with the angular velocity $\Omega_{{\rm p}, i}$ \cite{G&B2009}:
\begin{equation}
    \dot{\mathbf{S}}_i = \Omega_{{\rm p}, i} \times \mathbf{S}_{i} =\frac{G (4 + 3 q)}{2 c^2 r^3} \mathbf{L} \times \mathbf{S}_i \,.
    \label{Eq:S_i_dot_L}
\end{equation}
Since the orientation of the total angular momentum vector $\mathbf{J} = \mathbf{L} + \mathbf{S}_1 + \mathbf{S}_2$ is constant in this motion, the precession can be described around $\mathbf{J}$, with $\mathbf{L}$ also performing a precession around it. 
During the emission of GWs, the magnitudes of $\mathbf{L}$ and $\mathbf{J}$ are shrinking, while the angle $\alpha$ between $\mathbf{L}$ and $\mathbf{J}$ increases. 
At the same time, the angles $\beta_1$ between $\mathbf{S}_1$ and $\mathbf{J}$ and $\beta_2$ between $\mathbf{S}_2$ and $\mathbf{J}$ decrease, since $\alpha + \beta_1$ and $\alpha + \beta_2$ stay constant \cite{G&B2009}.
In the 2.5 post-Newtonian (PN) approximation, the direction of $\mathbf{J}$ stays constant, when the precessional angular velocity of the spins $\Omega_{{\rm p}, i}$ is larger than $\dot{\alpha}$, which is the case during the inspiral stage. 

At a later time $t_2$ during the inspiral stage, the angle $\alpha$ increased and the angles $\beta_1$ and $\beta_2$ decreased so much that, in this case, the spin $\mathbf{S}_1$ and thus the jet of the heavier black hole $m_1$ could point at Earth.
That is if Earth lies inside the opening angle of the jet, indicated by the orange area in Fig.~\ref{fig:Prediction_compare_2Spin}. 
Due to the precession of the jet around $\mathbf{J}$, the jet cone will move along the blue ring area and point at Earth after a time $\Delta t$. 
Since the angle $\beta_1$ decreases, the blue ring will get smaller with time, shortening the time between each potential signal received at Earth. 
This is until a time $t_3$ is reached, at which the Earth will be outside the blue area, so that the jet will no longer be able to point at Earth. 

The directional angle $\phi$ of the precessing jet from the BH with the spin $\mathbf{S}_1$ can be determined by integrating its precessional velocity and delivers: 
\begin{eqnarray}
    \phi(\Delta T_{\rm GW}, M, q, \alpha, \beta_1, \beta_2) 
    &=& 2 \, (4 + 3 q) \, \left(\frac{5\, c}{32}\right)^{\frac{3}{4}} \,
    \left( \frac{\eta}{G\, M} \right)^{\frac{1}{4}} \,
    \left( \Delta T_{\rm GW} \right)^{\frac{1}{4}} 
    \left[ q^{-1} \cos(\beta_1) + q \cos(\beta_2) \right] \notag \\
    && \ + \quad 
    \frac{4 (4 + 3 q)}{3}  \left(\frac{5}{32}\right)^{\frac{5}{8}} 
    c^{\frac{9}{8}} 
    \left( \frac{\eta}{G \, M} \right)^{\frac{3}{8}} 
    \left( \Delta T_{\rm GW} \right)^{\frac{3}{8}} \, \cos(\alpha) \notag \\
    && \ + \quad \psi(\tau ,M, q, \alpha, \beta_1, \beta_2) \,. 
    \label{Eq:phi_T_GW}
\end{eqnarray}
Here, $G$ is the gravitational constant, $M = m_1 + m_2$ the total mass of the SMBBH, $c$ the speed of light and $\eta = {q}/{(1+q)^2}$.
The remaining time until merger is $\Delta T_{\rm GW}$. 
It is defined as \cite{G&B2009}:
\begin{equation}
    \Delta T_{\rm GW} = \frac{5 \, G\,M}{32 \, c^3} \varepsilon^{-4} \, \eta^{-1} \,,
    \label{eq:time_till_merger}
\end{equation}
with the PN parameter $\varepsilon \approx \varv/c$. A value of $\varepsilon = 10^{-3}$ denotes the beginning of the inspiral stage, while the value $\varepsilon = 10^{-1}$ marks its end.
The factor $\psi(\tau ,M, q, \alpha, \beta_1, \beta_2)$ is an integration constant, which describes the initial direction of the jet at a time $\tau$ during the inspiral stage.

Since the jet points at Earth every $360^\circ \pm \zeta$, the following relationship can be established (see \cite{Kun2022}):
\begin{equation}
    \phi(\Delta T_{\rm GW}, M, q, \alpha, \beta_1, \beta_2) = \phi(\Delta T_{\rm GW} - P_{\rm jet}, M, q, \alpha, \beta_1, \beta_2) \pm \zeta \,.
    \label{Eq:Det_T_GW}
\end{equation}
The precession period $P_{\rm jet}$ denotes the time that has passed between two signals from the same source. 
This relation describes that the jet angle is the same at a time $\Delta T_{\rm GW}$ until merger and a later time $\Delta T_{\rm GW} - P_{\rm jet}$ until merger, with a jet cone of $\zeta$.
That way, for a given parameter combination of $M, q, \alpha, \beta_1, \beta_2, \zeta$ and a measured time between two flares from the same source $P_{\rm jet}$, the time until the merger $\Delta T_{\rm GW}$ of the source, in case it is an inspiral SMBBH, can be determined.

Exploiting the same relation, but with the time until merger $\Delta T_{\rm GW}$ now determined, Eq.~\ref{Eq:Det_T_GW} can be used to calculate the next, shorter period $P_{\rm jet, 2}$ between the second time that the jet pointed at Earth and the future third time it will point at Earth:
\begin{eqnarray}
    \phi(\Delta T_{\rm GW} , M, q, \alpha, \beta_1, \beta_2) 
    = \phi(\Delta T_{\rm GW} - P_{\rm jet} - P_{\rm jet, 2}, M, q, \alpha, \beta_1, \beta_2) \pm 2 \zeta \,.
    \label{Eq:Det_P_jet_2}
\end{eqnarray}
This way, a prediction can be made when the next flare should occur. 
This relation can be expanded until an $n$-th flare from the SMBBH.
The condition is that the binary did not merge until the flare and that Earth is still in the path of the jet, as can be seen in Fig.~\ref{fig:Prediction_compare_2Spin} at time $t_2$.

\section{Prediction of Neutrino Flares from TXS~0506+056 with the Baikal Neutrino}
For the prediction of future neutrino flares from TXS~0506+056, the time between the neutrino detections in 2014/2015 \cite{IceCube2014_2018} and 2017 \cite{IceCube2017_2018} with $P_{\rm jet, 1} = 2.78 \pm 0.15 \,\text{years}$ was taken as an input in the model.
This way, the occurrence of the 2022 neutrino flare was predicted in \cite{deBruijn2020} and confirmed in \cite{BeckerTjus2022}. 

Here, the expanded model is used to test whether the 2021 neutrino detected by the Baikal collaboration \cite{baikal2022} could also originate from TXS~0506+056 and be consistent with the jet precession model.
For that, the model is tested with a large parameter set: The total mass is varied between $7 \cdot 10^7 \,\text{M}_\odot$, $1 \cdot 10^8 \,\text{M}_\odot$, $3 \cdot 10^8 \,\text{M}_\odot$, $5 \cdot 10^8 \,\text{M}_\odot$ and $7 \cdot 10^8 \,\text{M}_\odot$ and the half-opening angle $\zeta$ between $3^\circ$ and $6^\circ$ in $0.1^\circ$ steps. As for the other angles, $\alpha$ is tested between $75^\circ$ and $90^\circ$, while $\beta_1$ and $\beta_2$ are assumed to lie between $0^\circ$ and $20^\circ$. 
The mass ratio $q$ is a set between $0.01$ and unity. 

Labeling the 2014/2015 neutrino detection as the first neutrino flare from TXS~0506+056 and the 2017 neutrino detection as the second, there are two possibilities for the Baikal neutrino flare: (i) either is it the third neutrino flare and the 2022 neutrino detection the 4th, or (ii) it is the 4th neutrino flare, the 2022 would be thus the 5th, with the third neutrino flare still hidden in the still-to-be analyzed IceCube data as suggested in \cite{BeckerTjus2022}.

\begin{figure}[h]
    \centering
    \includegraphics[angle=0,scale=0.5]{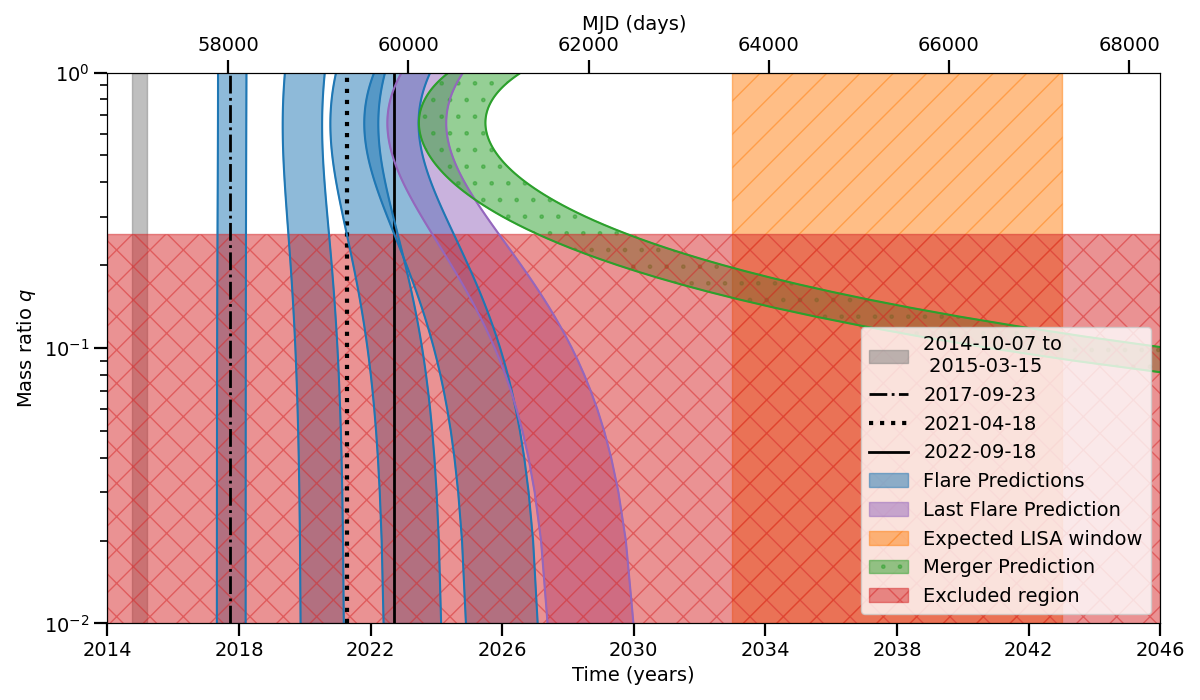}
    \caption{Prediction of the times for neutrino flares from TXS~0506+056 in case of an inspiral SMBBH close to its merger at its core and time of its merger in dependence of the mass ratio $q$. The assumption is that all four distinct neutrino flares detected originated from the blazar.}
    \label{fig:Baikal_pred}
\end{figure}

Performing a parameter study with the above mentioned parameters yields that there is no parameter combination for which (i) applies. 
The main reason is that the predicted time between the third and 4th flare is larger than the actual time between the 2021 and 2022 neutrinos. 
However, the situation is different for case (ii).
The parameter study showed that there are several cases possible for the Baikal neutrino to originate from this blazar in case of an inspiral SMBBH at its core.
The best-fit parameters are $M= 7 \cdot 10^8 \,\text{M}_\odot$, $\zeta = 4.6^\circ$, $\alpha = 89^\circ$ and $\beta_1 = \beta_2 = 20^\circ$ and $q=0.65$.
The prediction curves for these values in dependence of the mass ratio $q$ are shown in Fig.~\ref{fig:Baikal_pred}.
On the x-axis, the time in years and on the upper x-axis in MJD is shown, while the mass ratio is on the y-axis. 
The gray area marks the occurrence of the 2014/2015 neutrino flare, while the dashed-dotted, dotted and solid vertical lines show the time of the 2017, 2021 and 2022 neutrino flares, respectively. 
In blue, the predicted time bands for the next neutrino flares are shown. The prediction for the next neutrino flare is highlighted in purple for a better distinction.
The green area indicates the time band during which the actual merger of the binary will occur. 
In orange, the observational window of the Laser-interferometer Space Antenna (LISA) is drawn, expected to lie between 2033 and 2043. It should be sensible for a detection of GWs from SMBBH mergers until a total mass of $\sim 10^8 \,\text{M}_\odot$.
Finally, the red crossed area marks mass ratios for which the model does not work. 
That entails the condition $q \ge 0.26$ for the model to work.
This is because at smaller mass ratios, the 2022 neutrino flare lies outside the prediction bands for the 5th flare.

As can be seen in Fig.~\ref{fig:Baikal_pred}, all four neutrinos flares are in agreement with a jet precession origin if the mass ratio of the binary is above $q = 0.25$.
However, at such mass ratios, the binary could merge as early as in the year 2024 and in the year 2029/2030 at the latest.
This is way before the first observational run by LISA, so that no GW detection of the merging binary could be possible. 

Since before the merger the jet will most likely point at Earth again, a prediction on when the next neutrino flare might occur can be made. 
For the allowed mass ratios, in case all four neutrino flares originated from TXS~0506+056, the next neutrino flare is expected to be between November 2023 and November 2025 (purple are in Fig.~\ref{fig:Baikal_pred}).
In that case, the third, not yet detected neutrino flare, must have occurred between Mai 2019 and October 2020 and must be still hidden in the IceCube data.

\section{Prediction of Neutrino Flares from TXS~0506+056 without the Baikal Neutrino}
\begin{figure}[h]
    \centering
    \includegraphics[angle=0,scale=0.5]{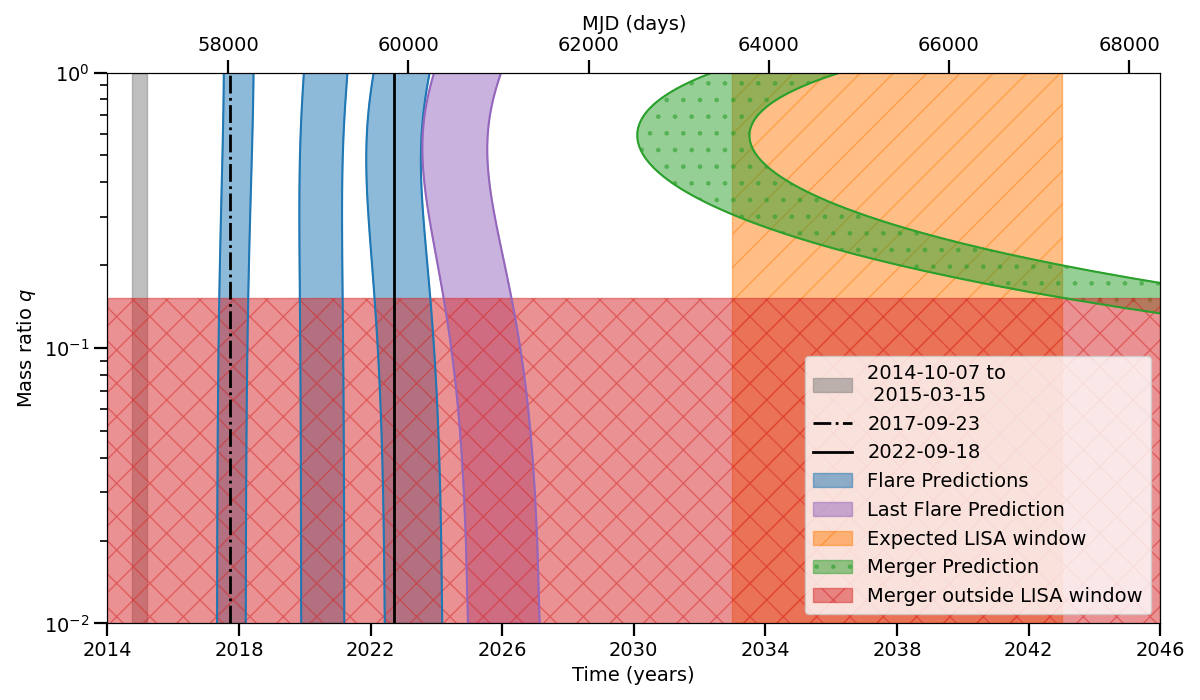}
    \caption{Prediction of the times for neutrino flares from TXS~0506+056 in case of an inspiral SMBBH close to its merger at its core and time of its merger in dependence of the mass ratio $q$. The assumption is that only the neutrinos detected by IceCube originated from the blazar.}
    \label{fig:IC_pred2}
\end{figure}

The same parameter study as described above has been performed only with the three neutrino flares detected by IceCube. 
Then, the 2022 neutrino will be the 4th neutrino from the source.
In this case, the best-fit parameters are $M= 3 \cdot 10^8 \,\text{M}_\odot$, $\zeta = 4.5^\circ$, $\alpha = 83^\circ$ and $\beta_1 = \beta_2 = 0^\circ$ and $q=0.45$.
The respective prediction curves in dependence of the mass ratio $q$ are illustrated in Fig.~\ref{fig:IC_pred2}. 
The descriptions and axes are the same as in Fig.~\ref{fig:Baikal_pred}, with the exception that the red area indicates the mass ratio for which the merger of the binary will not occur inside the expected LISA window. 

As is seen in Fig.~\ref{fig:IC_pred2}, all mass ratio agree with a currently ongoing SMBBH merger with this parameter combination.
However, LISA will be able to detect the merger of such a binary only if it has a mass ratio between $q=1$ and $q \approx 0.15$. And even then, there is the possibility that the merger happens outside the assumed LISA observational window, as can be seen in Fig.~\ref{fig:IC_pred2}. 

Nevertheless, for such a SMBBH scenario, a third neutrino flare must have happened between November 2019 and Mai 2021. 
The next neutrino flare is then expected to happen between July 2023 and March 2027. 



\section{Conclusions}
We expanded the analytical jet precession model, introduced in \cite{deBruijn2020}, by involving the second spin and the orbital angular momentum into the equations.
An application of the model on the three neutrino flares detected by IceCube from the direction of the blazar TXS~0506+056 and one neutrino detected by the Baikal Collaboration shows that all neutrino flares are consistent with a processing jet induced by an inspiral SMBBH close to its merger. 
However, the actual merger of the binary will occur before LISA is online and will thus not be detectable in GWs.

In case that only the IceCube neutrinos originated from the blazar, LISA will be able to detect GWs from the merger, if the mass ratio is between $1$ and $\sim 0.15$. 

In both cases, a neutrino flare should be still hidden in the not-analyzed IceCube data between 2019 and 2021, provided the conditions at the source are ideal for neutrino production.
The next neutrino flare should be happening between November 2023 and November 2025 if all four neutrino flares originated from the blazar and between July 2023 and March 2027 in case of only IceCube neutrinos originating from it. 
Again, the requirements are that the conditions for neutrino productions are fulfilled in the source environment.

It should be noted that although all four distinct neutrino flares investigated are consistent with the jet precession model, they are no confirmation of it. 
This is because the uncertainty regions of the 2021 Baikal neutrino and the 2022 IceCube neutrino are large ($\sim 6^\circ$ \cite{baikal2022} and $\sim 3.6^\circ$ \cite{Blaufuss2022}, respectively), so that the possibility remains that a different source than TXS~0506+056 is responsible for these two neutrino signals.

\begin{acknowledgments}
We acknowledge support from the Deutsche Forschungsgemeinschaft DFG, within the Collaborative Research Center SFB1491 "Cosmic Interacting Matters - From Source to Signal" (project No.\ 445052434) and from the project "MICRO" (project No. 445990517).
\end{acknowledgments}

%
%
%

\end{document}